\documentclass[a4paper]{PoS}

\title{Investigating the X-ray Emission from the Galactic TeV Gamma-ray Source MGRO~J1908+06}

\ShortTitle{Investigating the X-ray Emission from MGRO~J1908+06}

\author{\speaker{Dirk Pandel}\\
        Grand Valley State University, Allendale, MI 49401, USA\\
        E-mail: \email{pandeld@gvsu.edu}}

\abstract{
MGRO~J1908+06 is a bright, extended TeV gamma-ray source located near the Galactic plane.
The TeV emission has previously been attributed to the pulsar wind nebula
of the radio-faint gamma-ray pulsar PSR~J1907+0602 discovered with \textit{Fermi}.
However, studies of the TeV morphology with VERITAS have shown that MGRO~J1908+06
is somewhat larger than other pulsar wind nebulae of similar age and that the TeV spectrum
does not soften with distance from the pulsar as is observed for other pulsar wind nebulae.
Although MGRO~J1908+06 is very bright in gamma rays with a flux corresponding to $\sim$80\%
of the Crab Nebula flux at 20 TeV, no extended emission at other energies has so far been detected.
We report on our analysis of X-ray data obtained with \textit{XMM-Newton} of the region near MGRO~J1908+06.
We searched the data for point-like sources and detected several hard-spectrum X-ray sources
that could be associated with the TeV emission, including the gamma-ray pulsar PSR~J1907+0602.
We also performed an extended source analysis to search for diffuse emission from MGRO~J1908+06
but found no evidence of diffuse X-ray emission coincident with the TeV source.
We place an upper limit of $8.7\times10^{33}\rm~erg~s^{-1}$ on the X-ray luminosity
of MGRO~J1908+06 in the 1--10~keV energy range.
The corresponding limit on the ratio of gamma-ray to X-ray luminosity
is consistent with the ratios found for other pulsar wind nebulae of similar age.
}

\FullConference{The 34th International Cosmic Ray Conference\\
		30 July -- 6 August, 2015\\
		The Hague, The Netherlands}

\begin{document}

\section{Introduction}

In a survey of the Galactic plane at a median energy of 20~TeV,
the Milagro collaboration discovered the bright TeV gamma-ray source MGRO~J1908+06
with a flux of $\sim$80\% of the Crab Nebula flux at these energies \cite{2007ApJ...664L..91A}.
The source was subsequently detected with H.E.S.S.\ above 300~GeV
\cite{2008AIPC.1085..273D,2009A&A...499..723A} and with VERITAS \cite{2008AIPC.1085..301W}.
The H.E.S.S.\ observations showed that MGRO~J1908+06 (HESS~J1908+063) is an extended source,
and a fit of a two-dimensional Gaussian function to the excess map yielded an intrinsic size
of the TeV source of $\sigma=0.34^\circ$ and a centroid position of
$\rm{R.A.=19^h07^m54^s.3}$ and $\rm{Decl.=+06^\circ16'07''}$
($l=40^\circ23'9''$ and $b=-0^\circ47'10''$) \cite{2009A&A...499..723A}.
Although the TeV source is comparatively bright with a flux of 17\% of that of the Crab Nebula
above 1~TeV, no extended emission associated with MGRO~J1908+06 has so far been detected at other energies.
In particular, no extended GeV emission was found with the \textit{Fermi} Large Area Telescope (LAT),
which suggests that the spectrum of MGRO~J1908+06 has a low-energy turnover
between 20 and 300~GeV \cite{2011ApJ...726...35A}.

MGRO~J1908+06 may be associated with the nearby shell-type supernova remnant (SNR) G40.5--0.5.
However, the TeV emission extends well beyond the boundary of the radio SNR,
which would require either an additional source of gamma rays or the presence of dense molecular matter
interacting with ultra-relativistic particles in the vicinity of the SNR \cite{2009A&A...499..723A}.
A point-like GeV gamma-ray source, 0FGL J1907.5+0602, was discovered with
the \textit{Fermi} LAT near the position of MGRO~J1908+06 \cite{2009ApJS..183...46A}.
The source was later identified as a radio-quiet gamma-ray pulsar, PSR~J1907+0602,
with a pulse period of 106.6~ms, a spin-down power of $\sim$2.8$\times10^{36}\rm~erg~s^{-1}$,
and a characteristic age of 19.5~kyr \cite{2009Sci...325..840A,2010ApJ...711...64A}.
Using a time differencing technique, a precise position of the pulsar of
$\rm{R.A.=19^h07^m54^s.7(2)}$ and $\rm{Decl.=+06^\circ02'16(2)''}$
was derived from the \textit{Fermi} data.
Multi-wavelength observations of PSR~J1907+0602 revealed a faint X-ray counterpart
as well as faint, pulsed radio emission \cite{2010ApJ...711...64A}.
The dispersion measure obtained from the radio observations indicates a distance
of $3.2\pm0.6$~kpc to the pulsar.
The pulsar's position is offset by $\sim$0.23$^\circ$ from the center of HESS~J1908+063
but firmly within the extent of the TeV source.
The close proximity of the two sources suggests that the TeV emission
is the pulsar wind nebula (PWN) of PSR~J1907+0602.
The pulsar may also be associated with the SNR G40.5--0.5 \cite{2010ApJ...711...64A}.
The estimated age (20--40 kyr \cite{1980A&A....92...47D})
and distance (3.6 kpc \cite{2006ChJAA...6..210Y}) of the SNR
are consistent with those of the pulsar.
However, if the pulsar originated at the center of the SNR, its characteristic age would imply
a transverse velocity of $\sim$1400~${\rm km\,s^{-1}}$.
This is several times higher than typical pulsar velocities,
although similar velocities have been observed in a few cases \cite{2014ApJ...787..166A}.
At such a high pulsar velocity, a bow shock and pulsar tail will likely be present at X-ray or radio energies.
Evidence of a bow shock consistent with the pulsar moving away from the SNR was found
in X-ray images obtained with \textit{XMM-Newton} \cite{2012AIPC.1505..329P}.

Deep TeV observations of MGRO~J1908+06 were carried out with VERITAS
to investigate the morphology of the source \cite{2014ApJ...787..166A}.
The TeV emission was found to extend from the region near PSR~J1907+0602
to the boundary of the SNR G40.5--0.5.
However, in contrast to other PWNe of similar age, the TeV spectrum
does not soften with distance from the pulsar.
This is inconsistent with the TeV emission near the SNR being solely due to the PWN
of PSR~J1907+0602.
If the pulsar was born at the center of the SNR, the leptons injected into the PWN closer to
the pulsar's birthplace would have cooled significantly via synchrotron radiation and inverse Compton scattering,
leading to a softer TeV spectrum near the SNR \cite{2014ApJ...787..166A}.
It has therefore been suggested that the TeV emission near the SNR is caused by the interaction
of the pulsar wind with nearby molecular clouds or the SNR shock.
It is also possible that another PWN, associated with an as yet undetected pulsar along the line of sight,
contributes to the TeV emission.

We present an analysis of X-ray data obtained with \textit{XMM-Newton} \cite{2001A&A...365L...1J}
of the region near MGRO~J1908+06.
The \textit{XMM-Newton} observations and our analysis of the X-ray data are described in \S2.
In \S3 we discuss our search for point-like X-ray sources that may be associated with the TeV source,
and in \S4 we investigate the diffuse X-ray emission from MGRO~J1908+06.
Our findings are discussed in \S5.

\section{Observations and Analysis}

MGRO~J1908+06 was observed with \textit{XMM-Newton} during four observations
with exposure times ranging from 7 to 18~ks
(Observation IDs 0553640101, 0553640201, 0553640701, and 0553640801).
Because of the large extent of the TeV source, the observations were arranged
in a 2-by-2 mosaic covering a field $0.75^\circ$ across with the X-ray cameras.
We analyzed the data using the \textit{XMM-Newton} Science Analysis Software (SAS)
and included in our analysis data from the two EPIC MOS cameras \cite{2001A&A...365L..27T}
and the EPIC PN camera \cite{2001A&A...365L..18S}.

To search for point-like X-ray sources, we used the SAS task \textit{emosaicproc}
with the combined data from the three cameras and four observations
in the 0.4--10~keV energy range.
We detected 22 sources with a likelihood $DET\_ML>15$
which corresponds to a probability of $<$3$\times10^{-7}$
for the observed source counts to be caused by a background fluctuation.
For each of the detected sources, we extracted a background-subtracted spectrum
using a $20''$ circular aperture and determined the hardness ratio
$HR=(F_{2-10~keV}-F_{0.5-2~keV})/F_{0.5-10~keV}$ of the observed fluxes
in the 0.5--2~keV and 2--10~keV energy ranges.

\begin{figure}[!t]
\centering
\includegraphics[width=1.0\textwidth]{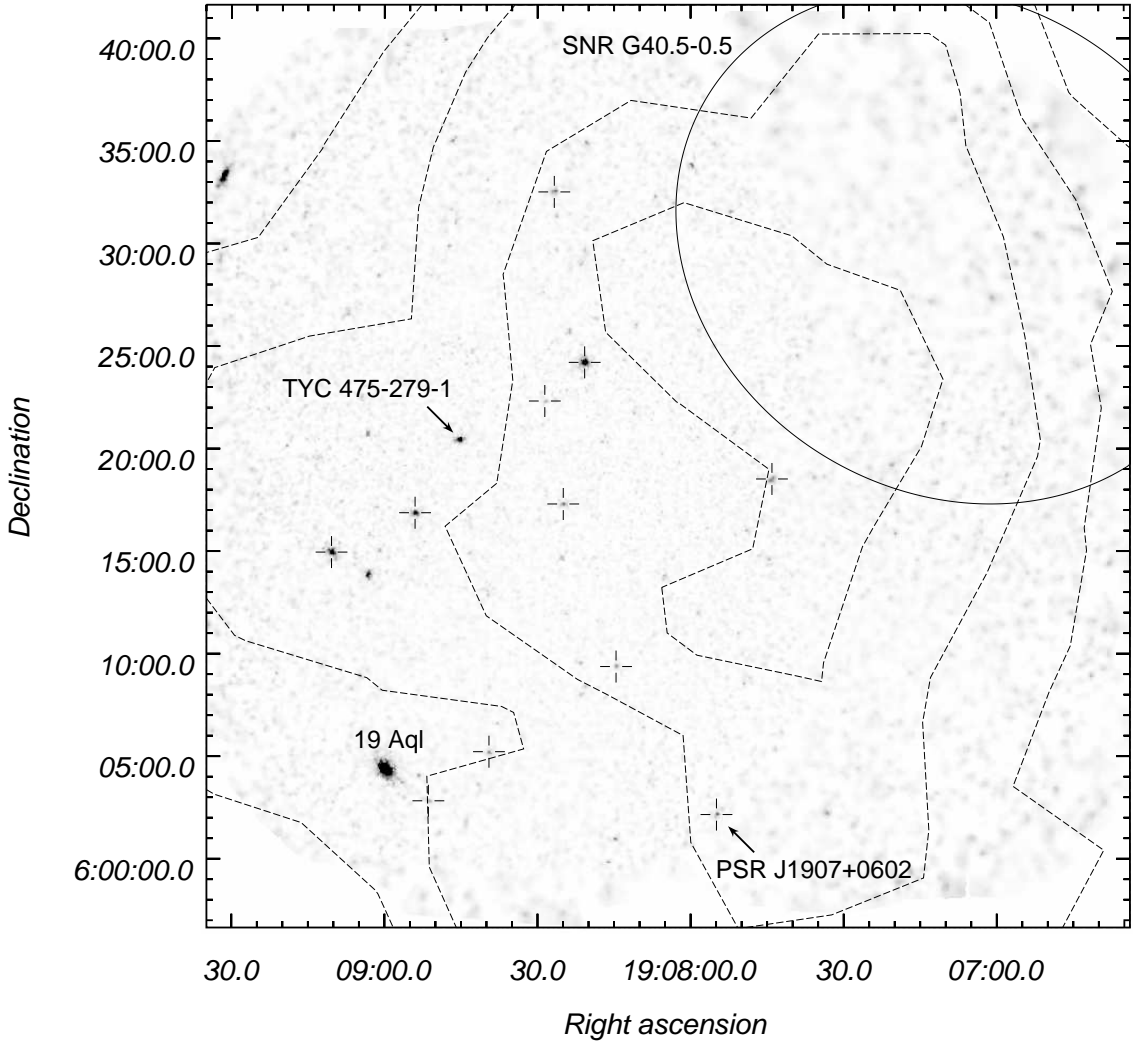}
\caption{
X-ray image of the region near MGRO~J1908+06 in the 0.4--7.2~keV energy range.
The image is $0.75^\circ$ across and shows the combined data from the three X-ray cameras
and four observations.
The image has been exposure corrected, background subtracted (except for the residual soft proton background
and some instrumental lines between 1.2 and 2.0~keV), and adaptively smoothed.
The crosses indicate the positions of the detected X-ray sources with a hard spectrum (hardness ratio $HR>0$).
The dashed contours show a significance map (7$\sigma$, 6$\sigma$, 5$\sigma$, and 4$\sigma$ levels)
of the TeV gamma-ray emission detected with H.E.S.S.\ (from \cite{2008AIPC.1085..273D}).
The ellipse outlines the extent of the shell-type SNR G40.5--0.5.
Note that the effective exposure time varies significantly across the image, leading to an uneven
appearance of the diffuse emission after the adaptive smoothing is applied.
}
\label{xrayimage}
\end{figure}

To investigate the diffuse X-ray emission from MGRO~J1908+06, we analyzed the data
using the Extended Source Analysis
Software\footnote{ftp://legacy.gsfc.nasa.gov/xmm/software/xmm-esas/xmm-esas.pdf} (ESAS).
We excluded from our analysis time intervals with strong contamination by soft proton flares
and EPIC MOS CCDs that were operating in an anomalous state.
Using the ESAS, we modeled the particle background for each observation and camera
and created background-subtracted images and spectra while taking into account
the varying effective exposure times across the field of view.
An image of the combined X-ray data from the three cameras and four observations
is shown in Figure~\ref{xrayimage}.
To obtain spectra of the diffuse X-ray emission for selected regions, we excluded all detected
point sources and combined the remaining data from all observations for each camera.
Because of the large extent of MGRO~J1908+06, it was not possible to estimate
the residual soft proton background from a region away from the TeV source.
We therefore did not subtract this background directly, but instead included it as a separate
model component when fitting the spectrum.
Our spectral analysis was restricted to the energy range 0.4--10~keV for the EPIC MOS cameras
and 0.4--12~keV for the EPIC PN camera.
To avoid strong emission lines in the instrumental background that are difficult to model,
we also excluded the 1.2--2.0~keV energy range for all cameras
and the 7.2--9.2~keV energy range for the EPIC PN camera.
The spectral analysis was performed using XSPEC \cite{1996ASPC..101...17A}.

\section{X-ray Point Sources}

Of the 22 point-like X-ray sources we detected, three are previously known sources observed
at other wavelengths: two foreground stars with a soft X-ray spectrum, 19~Aql and TYC 475-279-1,
and the gamma-ray pulsar PSR~J1907+0602.
The X-ray spectrum we found for PSR~J1907+0602 is consistent with that obtained from
\textit{Chandra} observations \cite{2010ApJ...711...64A}.
Sources with a hard X-ray spectrum (hardness ratio $HR>0$) are marked with crosses
in Figure~\ref{xrayimage}.
Note that the effective exposure time on the right side of the figure is $\sim$2.5 times lower
than on the left side, which may explain the larger number of detected sources on the left.

A possible explanation for the TeV emission away from PSR~J1907+0602
(near the left edge of SNR G40.5--0.5 in Figure~\ref{xrayimage})
is the presence of a second PWN associated with an as yet undetected pulsar
along the line of sight \cite{2014ApJ...787..166A}.
We detected several sources with a hard X-ray spectrum in this region which are potential candidates
for such a pulsar.
However, further multi-wavelength observations are necessary to determine if any of these sources
are pulsars or compact X-ray PWNe.

\section{Diffuse X-ray Emission}

\begin{figure}[!t]
\centering
\includegraphics[width=0.48\textwidth]{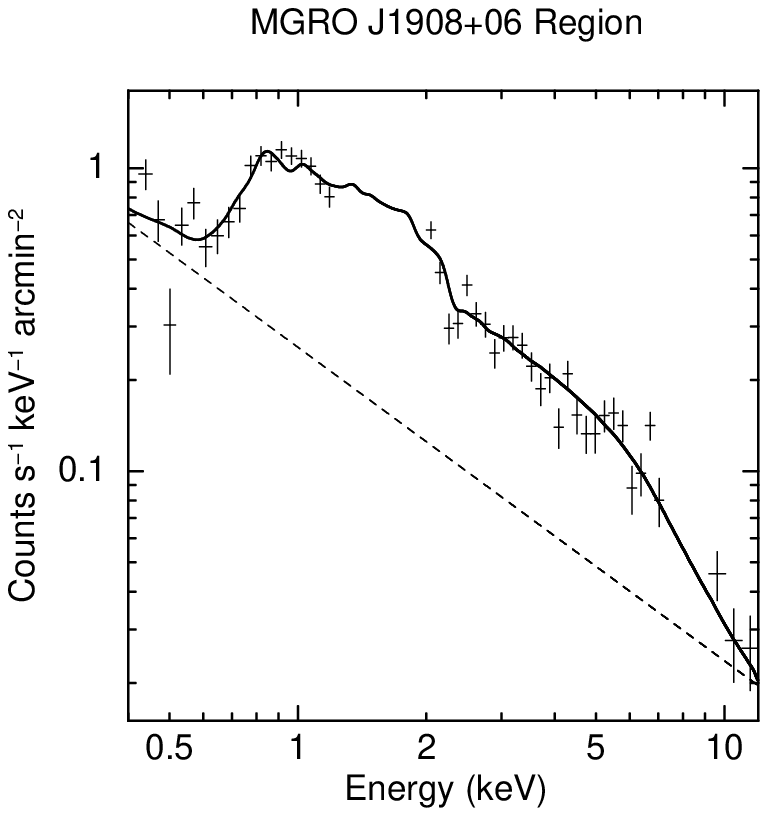}
\hfill
\includegraphics[width=0.48\textwidth]{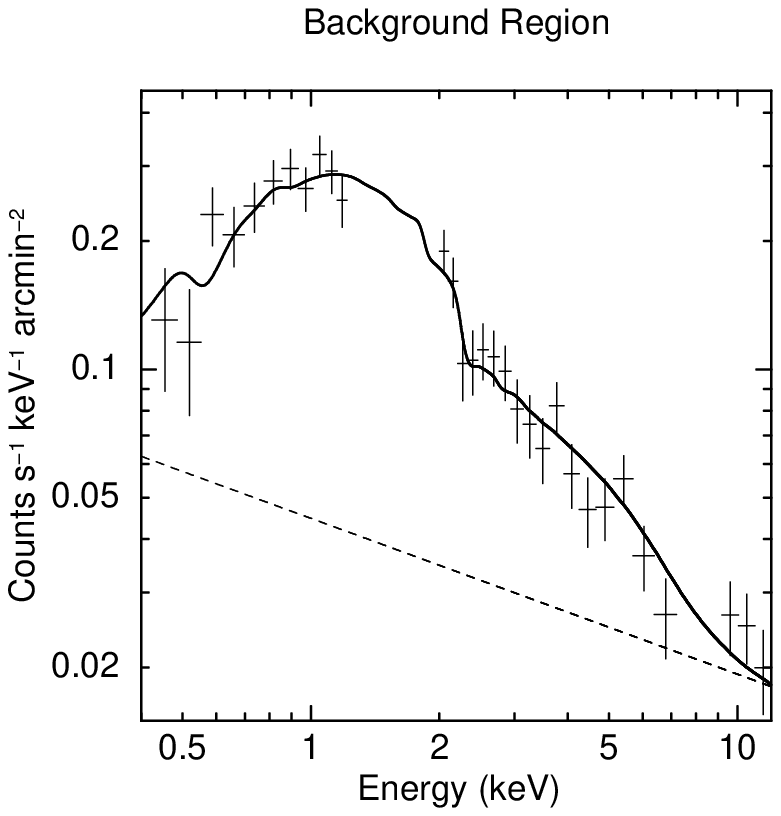}
\caption{
Spectrum of the diffuse X-ray emission from a $45'$ circular region around MGRO~J1908+06
and from a background region (see \S4).
The data points show the particle background subtracted count rates measured with the EPIC PN camera
normalized to an area of one square arcminute.
The solid curve is a best fit with an absorbed \textit{mekal} plus power law model (left)
or an absorbed power law model (right).
The dashed line shows an estimate of the residual soft proton background obtained from the fit.
}
\label{xrayspectrum}
\end{figure}

Although some diffuse X-ray emission not associated with the point-like sources
is visible in Figure~\ref{xrayimage}, it appears to be uniform and does not show
any excess correlated with the TeV emission (dashed contours).
To investigate the spectrum of the diffuse X-ray emission near MGRO~J1908+06
we excluded the detected point sources and selected all remaining photons
from a circular region with a radius of $45'$ centered on
$\rm{R.A.=19^h07^m22^s}$ and $\rm{Decl.=+06^\circ13'12''}$,
the centroid position of the TeV emission determined with VERITAS \cite{2014ApJ...787..166A}.
To obtain a spectrum of the diffuse background emission,
we selected all photons outside the $45'$ circular region.
Figure~\ref{xrayspectrum} shows the count rate spectra from the EPIC PN camera
for the MGRO~J1908+06 region and the background region normalized to an area of one square arcminute.

We fitted the spectrum of the MGRO~J1908+06 region with a model for a hot diffuse gas (\textit{mekal})
plus a power-law model and included photoelectric absorption by the interstellar medium.
The spectrum of the background region was fitted with an absorbed power law model.
The best fits with these models are shown in Figure~\ref{xrayspectrum} as solid curves.
Although we were able to estimate and subtract most of the instrumental background,
the spectra still contain a residual contribution from the soft proton background.
We modeled this background separately for each camera as a power law
that is not folded through the instrumental response and included it
as an additional model component when fitting the spectrum (dashed lines in Figure~\ref{xrayspectrum}).

From the spectral fits we estimated the average energy flux and photon flux per square arc\-minute
for the two regions in the 1--10~keV energy range.
Table~\ref{fluxtable} shows the absorbed flux observed with \textit{XMM-Newton}
as well as the unabsorbed flux obtained after removing the effects of interstellar absorption.
The X-ray flux in the MGRO~J1908+06 region is consistent within uncertainties
with the flux in the background region.
We therefore find no evidence of any X-ray emission coincident with the TeV emission
from MGRO~J1908+06.
From the flux measurements we derive an upper limit of $7.1\times10^{-12}\rm~erg~cm^{-1}~s^{-1}$
or $1.4\times10^{-3}\rm~photons~cm^{-1}~s^{-1}$ (95\% confidence level)
on the 1--10~keV X-ray flux from MGRO~J1908+06.

\begin{table}
\begin{tabular}{lcc}
\hline
& MGRO~J1908+06 Region & Background Region \\
\hline
Flux in $10^{-14}\rm~erg~cm^{-1}~s^{-1}~arcmin^{-2}$ & & \\
~~~Absorbed     & $1.05\pm0.11$ & $1.12\pm0.17$ \\
~~~Unabsorbed & $1.18\pm0.10$ & $1.18\pm0.16$ \\
Flux in $10^{-6}\rm~photons~cm^{-1}~s^{-1}~arcmin^{-2}$ & & \\
~~~Absorbed     & $1.82\pm0.12$ & $2.02\pm0.22$ \\
~~~Unabsorbed & $2.38\pm0.20$ & $2.27\pm0.21$ \\
\hline
\end{tabular}
\caption{Absorbed and unabsorbed X-ray flux per square arcminute in the 1--10~keV energy range
from a $45'$ circular region around MGRO~J1908+06 and from a background region (see \S4).
Uncertainties are given at the 68\% confidence level.}
\label{fluxtable}
\end{table}

\section{Discussion}

According to current models, the emission from PWNe is produced by a population of leptons
accelerated to very high energies at the shocks formed by the pulsar wind interacting with the ambient medium.
The spectrum of the emission has two broad components,
one at X-ray to radio energies generally understood as synchrotron radiation from the highly energetic leptons,
and one at gamma-ray energies caused by Inverse Compton (IC) scattering of ambient low-energy photons.
Because of the spin down of the pulsar and the cooling of the leptons,
both spectral components decrease with the age of the PWN.
However, the decaying magnetic field causes the synchrotron component to decrease more rapidly
than the IC component \cite{2010ApJ...715.1248T}.
In agreement with this model, the data compiled by \cite{2013arXiv1305.2552K}  on known PWNe
show a positive correlation between the gamma-ray to X-ray luminosity ratio $L_\gamma/L_X$
and the pulsar age.

Our upper limit on the 1--10~keV X-ray flux from MGRO~J1908+06
corresponds to a luminosity of $L_X\le8.7\times10^{33}\rm~erg~s^{-1}$
for a distance of 3.2~kpc.
According to Fig.~2 in \cite{2013arXiv1305.2552K},
this upper limit is 1--2 orders of magnitude higher than the X-ray luminosity
of most PWNe containing a pulsar with a similar spin-down power as PSR~J1907+0602
($\sim$2.8$\times10^{36}\rm~erg~s^{-1}$).
The gamma-ray luminosity of MGRO~J1908+06 is $L_\gamma=3\times10^{34}\rm~erg~s^{-1}$
in the 1--10 TeV energy range \cite{2014ApJ...787..166A}.
This implies a lower limit on the gamma-ray to X-ray luminosity ratio of $L_\gamma/L_X\gtrsim3$.
For other PWNe with a similar age as MGRO~J1908+06 ($\sim$19.5~kyr), this ratio is typically in the 1--500 range
(Fig.~4 in \cite{2013arXiv1305.2552K}).
We therefore conclude that the non-detection of X-ray emission from MGRO~J1908+06
is consistent with the X-ray luminosities found for other PWNe with a similar pulsar age
and spin-down power as PSR~J1907+0602.

\begin{figure}[!t]
\centering
\includegraphics[width=0.8\textwidth]{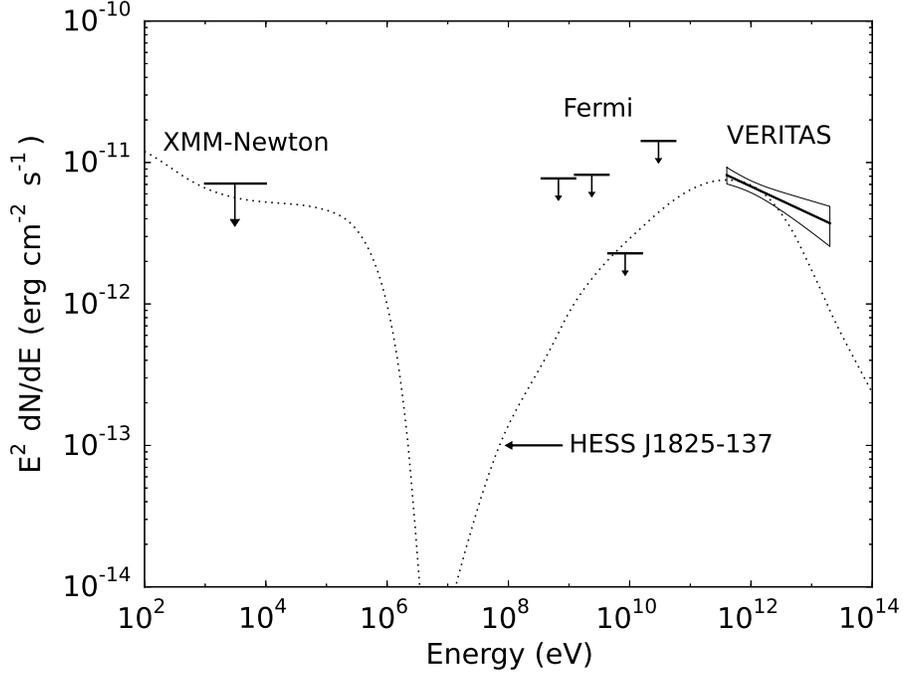}
\caption{
TeV spectrum of MGRO~J1908+06 obtained with VERITAS \cite{2014ApJ...787..166A}
shown together with the \textit{Fermi} upper limits on the off-pulse GeV emission
from the pulsar region \cite{2010ApJ...711...64A} and our upper limit on the X-ray flux.
For comparison, the dotted curve shows a modeled spectral energy distribution of the PWN \mbox{HESS J1825--137}
scaled to match the TeV flux of MGRO~J1908+06 (from \cite{2012arXiv1202.1455M}).
}
\label{sed}
\end{figure}

Time-dependent models that take into account the injection and cooling of leptons in PWNe
have been used to investigate the evolution of the spectral energy distribution from radio to TeV energies
\cite{2010ApJ...715.1248T,2012arXiv1202.1455M}.
Such models could be used to predict the X-ray luminosity of MGRO~J1908+06
from the observed TeV spectrum or to constrain certain PWN parameters
such as the magnetic field strength.
Figure~\ref{sed} shows the TeV spectrum of MGRO~J1908+06 together with our upper limit on the X-ray flux
and the \textit{Fermi} upper limits on the off-pulse GeV emission \cite{2010ApJ...711...64A}.
For comparison, the figure also shows a modeled spectral energy distribution of the PWN HESS~J1825--137
(from \cite{2012arXiv1202.1455M}) which has a similar age and pulsar spin-down power as MGRO~J1908+06.

\acknowledgments

This work is based on observations obtained with \textit{XMM-Newton}, an ESA science mission
with instruments and contributions directly funded by ESA member states and the USA (NASA).
This material is based upon work supported by NASA Guest Investigator Grant NAG57816
and by the National Science Foundation under Grant No.\ 1068152.

\end{document}